\documentclass[referee]{aa}
\usepackage{graphicx}
\usepackage[varg]{txfonts}
\usepackage{cases}

\begin{document}
   \title{Blobs in recurring EUV jets}

   \author{Q. M. Zhang \and H. S. Ji}

   \institute{Key Laboratory for Dark Matter and Space Science, Purple
              Mountain Observatory, CAS, Nanjing 210008, China \\
              \email{zhangqm@pmo.ac.cn}
              }

   \date{Received; accepted}
    \titlerunning{Blobs in EUV jets}
    \authorrunning{Q. M. Zhang \& H. S. Ji}

  \abstract
   {Coronal jets are one type of ubiquitous small-scale activities caused by 
   magnetic reconnection in the solar corona. They are often associated with 
   cool surges in the chromosphere.}
   {In this paper, we report our discovery of blobs in the recurrent 
   and homologous jets that occurred at the western edge of NOAA active 
   region 11259 on 2011 July 22.}
   {The jets were observed in the seven extreme-ultraviolet (EUV) 
   filters of the Atmospheric Imaging Assembly (AIA) instrument 
   aboard the Solar Dynamics Observatory (SDO). Using the base-difference 
   images of the six filters (94, 131, 171, 211, 193, and 335 {\AA}), we 
   carried out the differential emission measure (DEM) analyses to explore 
   the thermodynamic evolutions of the jets. The jets were 
   accompanied by cool surges observed in the H$\alpha$ line center of
   the ground-based telescope in the Big Bear Solar Observatory.}
   {The jets that had lifetimes of 20$-$30 min recurred at the same place  
   for three times with interval of 40$-$45 min. Interestingly,
   each of the jets intermittently experienced several upward eruptions at the 
   speed of 120$-$450 km s$^{-1}$. After reaching the maximum heights, 
   they returned back to the solar surface, showing near-parabolic trajectories.
   The falling phases were more evident in the low-$T$ filters than in 
   the high-$T$ filters, indicating that the jets experienced cooling after the 
   onset of eruptions. We identified bright and compact blobs in the 
   jets during their rising phases. The simultaneous 
   presences of blobs in all the EUV filters were consistent with the broad 
   ranges of the DEM profiles of the blobs ($5.5\le \log T\le7.5$), indicating 
   their multi-thermal nature. The median temperatures of the blobs were 
   $\sim$2.3 MK. The blobs that were $\sim$3 Mm in diameter had lifetimes 
   of 24$-$60 s.}
   {To our knowledge, this is the first report of blobs in coronal jets. We 
   propose that these blobs are plasmoids created by the magnetic reconnection 
   as a result of tearing-mode instability and ejected out along the jets.}


   \maketitle

\section{Introduction} \label{s-intro}

Coronal jets are transitory X-ray or extreme-ultraviolet (EUV) enhancements 
with an apparent collimated motion (Shibata et al. \cite{shi92}). They were
first observed by the Soft X-ray Telescope (SXT) aboard Yohkoh (Tsuneta 
et al. \cite{tsu91}). More and more coronal jets were observed by space-borne 
telescopes with higher resolutions and time cadences in the last two decades
(Chae et al. \cite{chae99}; Savcheva et al. \cite{sav07}; Chifor et al. \cite{chi08};
Nistic{\`o} et al. \cite{nis09}; Shen et al. \cite{shen12}; 
Moschou et al. \cite{mos12}; Lee et al. \cite{lee13}; Jiang et al. \cite{jiang13}). 
It is widely accepted that coronal jets are heated by magnetic reconnection 
between emerging flux and the pre-existing magnetic fields with opposite polarity 
(Shibata \& Uchida \cite{shi86}; Yokoyama \& Shibata \cite{yoko96}; 
Moreno-Insertis et al. \cite{mor08}; T{\"o}r{\"o}k et al. \cite{tor09}; 
Moreno-Insertis \& Galsgaard \cite{mor13}). According to the numerical 
simulations of Yokoyama \& Shibata (\cite{yoko96}), there are hot, compact 
microflares at their footpoints (Krucker et al. \cite{kru11}; Zhang \& Ji \cite{zqm13}) 
and cool H$\alpha$ surges adjacent to the hot jets (Schmieder et al. \cite{sch95}; 
Canfield et al. \cite{can96}; Liu \& Kurokawa \cite{liu04}; Jiang et al. \cite{jiang07}; 
Liu \cite{liu08}; Nelson \& Doyle \cite{nel13}). The EUV emissions from a hot jet 
were observed to be absorbed by the foreground cool surge, resulting in EUV 
dimming behind the leading edge of the jet (Zhang \& Ji \cite{zqm14}).
Pariat et al. (\cite{pari09}) proposed a new mechanism for the coronal 
hole jets as a result of continuous pumping of magnetic free energy as well 
as helicity into the upper solar atmosphere. The presence of dome-like magnetic 
topology that consists of a null point, a spine, and a separatrix surface
in a X-ray bright point associated with recurrent jets was observed and 
reported by Zhang et al. (\cite{zqm12}). Coronal jets are usually generated
in coronal holes (Cirtain et al. \cite{cir07}; Culhane \cite{cul07}; 
Patsourakos et al. \cite{pat08}; Chandrashekhar et al. \cite{chan14}) 
or at the edge of active regions (Kim et al. \cite{kim07}; Guo et al. \cite{guo13}; 
Schmieder \cite{sch13}; Zhang \& Ji \cite{zqm14}). 
The apparent heights and widths of jets are 10$-$400 Mm and 5$-$100 Mm.
The velocities of jets are 10$-$1000 km s$^{-1}$, which are the same order of 
magnitude as the coronal Alfv\'{e}n speed (Shimojo et al. \cite{sho96}).

Recurrent jets and surges have often been observed and extensively been studied 
thanks to the everlasting development and improvement of the solar telescopes. 
Chae et al. (\cite{chae99}) analysed simultaneous EUV data and H$\alpha$ data from BBSO.
Several EUV jets repeatedly occurred in the active region where pre-existing magnetic 
flux was canceled by newly emerging flux of opposite polarity. 
Chifor et al. (\cite{chi08}) analysed a recurring solar active region jet observed in X-ray 
and EUV, finding a correlation between recurring magnetic cancellation and the X-ray 
jet emission. The jet emission was attributed to chromospheric evaporation flows due 
to recurring magnetic reconnection. Recurrent jets can also be caused by moving 
magnetic features (Brooks et al. \cite{bro07}; Yang et al. \cite{yang13}).
Guo et al. (\cite{guo13}) reported the discovery of three EUV jets recurring in about 
one hour on 17 September 2010. According to the nonlinear force-free field extrapolation,  
the authors concluded that the magnetic reconnection occurred periodically in the current layer 
created between the emerging bipoles and the large-scale active region field, inducing the 
observed recurrent coronal jets. Murray et al. (\cite{mur09}) performed MHD numerical 
simulations of interaction between emerging flux and pre-existing magnetic fields in a 
coronal hole. When gas pressure of the reconnection outflow region exceeds that of 
the inflow region, the magnetic field lines in the two bounded outflow regions are driven 
to reconnect reversely, giving rise to an oscillatory reconnection and recurrent jets with 
the peak magnetic reconnection rate decreasing as time goes on. 
Pariat et al. (\cite{pari10}) performed 3D MHD simulations of periodic coronal jets due 
to continuous twisting motion of the photosphere and pumping of magnetic free energy. 

The temperatures of coronal jets have been extensively investigated since their 
discovery. There are mainly two methods for temperature diagnostics. One 
is filter-ratio assuming single temperature along the line-of-sight (LOS). It 
has been applied using the EUV 171 {\AA}, 195 {\AA}, and 284 {\AA} filters 
or broad-band soft X-ray filters (e.g., Shimojo \& Shibata \cite{sho00}; 
Nistic{\`o} et al. \cite{nis11}; Madjarska \cite{mad11}; Madjarska et al. \cite{mad12}; 
Matsui et al. \cite{mat12}; Pucci et al. \cite{puc13}; Young \& Muglach \cite{you13}). 
This approach, however, has its limitation and weakness. The reliable temperatures 
are limited within the monotonic range of the ratio of the temperature response 
functions of the two employed filters. The calculated temperature may represent 
the average temperature of the plasmas along the LOS. The other method is differential 
emission measure (DEM) based on the multi-thermal nature of the plasmas along the 
LOS (Doschek et al. \cite{dos10}; Chandrashekhar \cite{chan13}; Kayshap et al. \cite{kay13}; 
Chen et al. \cite{chen13}; Sun et al. \cite{sun14b}), which is more realistic and reasonable.
Occasionally, hard X-ray (HXR) sources could be detected at the bottom of jets, 
and the hot thermal component could be isolated during the fittings to perform 
temperature diagnostic (Bain \& Fletcher \cite{bain09}; Krucker et al. \cite{kru11}; 
Glesener et al. \cite{gle12}). The temperatures of jets are 0.5$-$8 MK in the main 
body and even higher (10$-$30 MK) at the bottom, which 
have been reproduced in the multi-dimensional magnetohydrodynamic (MHD) 
numerical simulations (Nishizuka \cite{nis08}; Archontis \& Hood \cite{arc13}).

Blob-like features or plasmoids are ubiquitous in the solar atmosphere.
During the magnetic reconnections involved in solar flares, the electric current sheet may 
subject to the tearing-mode instability (Furth et al. \cite{fur63}), leading to the formation
of multiple magnetic islands or plasmoids that are bidirectionally ejected out of the diffusion 
region (Ohyama \& Shibata \cite{ohy98}; Kliem et al. \cite{kli00}; Ko et al. \cite{ko03}; 
Asai et al. \cite{asa04}; 
Lin et al. \cite{lin05}; Ko{\l}oma{\'n}ski \& Karlick{\'y} \cite{ko07}; B{\'a}rta et al. \cite{bar08}; 
Nishizuka et al. \cite{nis10}; Milligan et al. \cite{mil10}; Takasao et al. \cite{tak12}; 
Ni et al. \cite{ni12a,ni12b}; Kumar \& Cho \cite{kum13}). The plasmoid velocity is found 
to have a positive correlation with the reconnection rate (Nishida et al. \cite{nish09}).
White-light blobs 
are observed to be quasi-periodically ejected out of the large-scale coronal streamers  
(Song et al \cite{song09}). Recurrent plasmoids in the chromospheric anemone jets with 
size of $\sim$0.1 Mm are observed by Singh et al. (\cite{singh12}) and reproduced in the 
numerical simulations of Yang et al. (\cite{yang13}). In this 
paper, we investigated the thermodynamic evolutions of the recurrent and homologous jets 
observed by the Atmospheric Imaging Assembly (AIA; Lemen et al. \cite{lem12}) aboard the 
Solar Dynamics Observatory (SDO) on 2011 July 22 and report our discovery of blobs in 
the jets. In Sect.~\ref{s-data}, we describe the multi-wavelength data analysis followed by the 
results in Sect.~\ref{s-result}. Discussions and summary are presented in Sect.~\ref{s-disc} 
and Sect.~\ref{s-sum}, respectively.

\section{Data analysis} \label{s-data}

There are seven EUV filters (94, 131, 171, 193, 211, 304, and 335 {\AA}) aboard the 
SDO/AIA instrument to achieve a wide temperature coverage ($5.5\le \log T \le7.5$). 
On 2011 July 22, the homologous jets recurred for three times at the western edge of 
NOAA active region (AR) 11259 during 21:00$-$23:30 UT. They were observed by 
AIA in all the EUV wavelengths with time cadence of 12 s and resolution of 
1$\farcs$2. The level\_1 fits data were calibrated using the standard program 
{\it aia\_prep.pro} in the Solar Software. The images observed in different wavelengths 
were coaligned carefully using the cross-correlation method. Fortunately,
the hot EUV jets were accompanied by cool H$\alpha$ surges observed by the 
ground-based telescopes in Big Bear Solar Observatory (BBSO) with time cadence 
of 85$-$100 s and resolution of 2$\arcsec$. The H$\alpha$ images were coaligned 
with the 304 {\AA} images.

The observed intensity of an optically thin EUV line $i$ is defined as 
$I_{i}=\int_{T_1}^{T_2}(d\mathrm{EM}/dT)R_{i}(T)dT$, 
where $\log T_1=5.5$ and $\log T_2=7.5$ stand for 
the minimum and maximum temperatures for the integral, $R_{i}(T)$ denotes the temperature 
response function of line $i$, and ${\mathrm{EM}}=\int n_{e}^{2}dh$ represents the total 
column emission measure along the LOS. Here, $n_{e}$ stands for the electron number 
density. Therefore, $\mathrm{DEM}=d\mathrm{EM}/dT=n_{e}^{2}dh/dT$ means the differential 
emission measure. The DEM-weighted average temperature of the plasma along the LOS 
$T_{eff}=\int_{T_1}^{T_2}\mathrm{DEM}\times T\times dT/\int_{T_1}^{T_2}\mathrm{DEM}dT
=\int_{T_1}^{T_2}\mathrm{DEM}\times T\times dT/\mathrm{EM}$. A couple of methods 
and programs have been developed to reconstruct the DEM profiles using a set of EUV 
or SXR filters (e.g., Golub et al. \cite{gol04}; Weber et al. \cite{web04,web05}; 
Hannah \& Kontar \cite{han12}; Aschwanden et al. \cite{asch13}). Using six of the AIA filters 
(94, 131, 171, 211, 193, and 335 {\AA}), Cheng et al. (\cite{cheng12}) derived the DEM profiles 
and average temperatures of the three components of a coronal mass ejection (CME; 
Chen \cite{chen11}) observed by SDO: hot channel in the core region, the bright loop-like leading 
front, and coronal dimming in the wake of the CME (Cheng et al. \cite{cheng13}; 
Hannah \& Kontar \cite{han13}). Recently, this method has also been applied to the 
temperature diagnostics of solar flares (Sun et al. \cite{sun14a}) as well as failed filament 
eruption (Song et al. \cite{song14}). Since the jets were quite close to the active region,
we selected the images before the onsets of the jets as base images and derived the 
base-difference images during the jets so that the emissions and influences of the 
background corona were removed. We used the base difference images and the same 
method to reconstruct the DEM profiles and study the thermodynamic evolutions of the 
recurring jets. To evaluate the confidence in the reconstructed DEMs, we conducted Monte 
Carlo (MC) simulations described in detail in Cheng et al. (\cite{cheng12}). The chi-square 
($\chi^2$) of the 100 MC simulations is a measure of scatter of the DEM profiles. 
Lower values of $\chi^2$ mean smaller uncertainties in the DEM solution.

\section{Results} \label{s-result}

\subsection{Recurrent EUV jets} \label{s-euv}

\begin{figure}
\centering
\includegraphics[width=10cm,clip=]{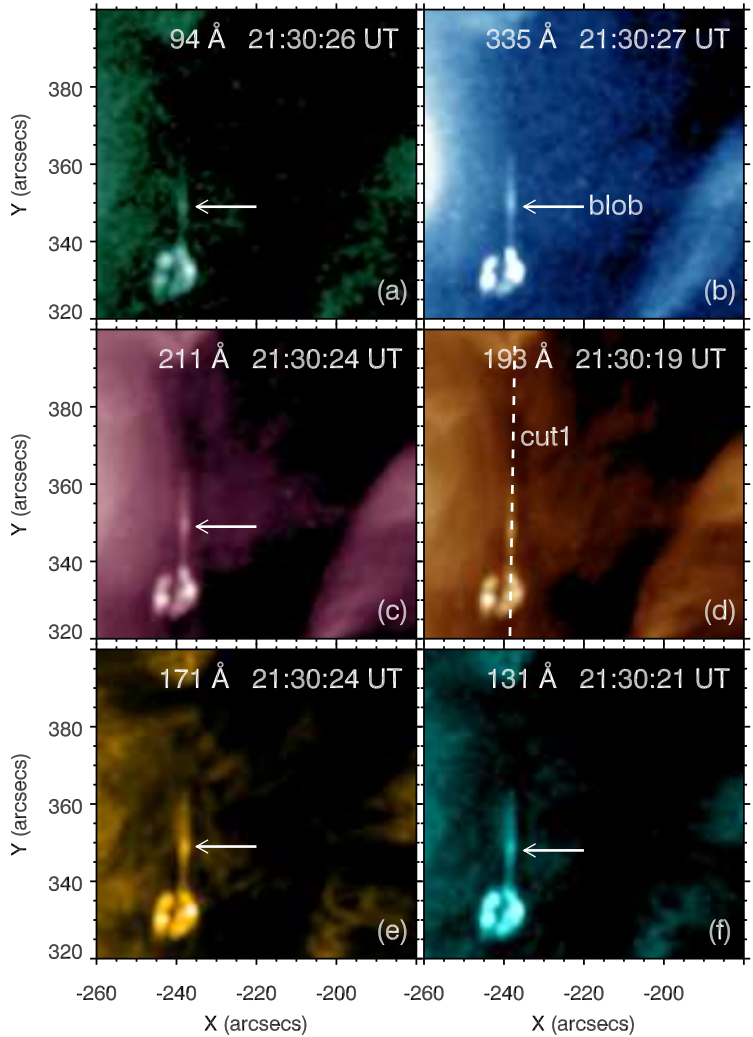}
\caption{Snapshots of jet1 seen in the six EUV filters at $\sim$21:30:20 UT. 
The white arrows point to the blobs within the jet in panels {\bf (a)}, {\bf (b)}, {\bf (c)}, 
{\bf (e)}, and {\bf (f)}. The white dashed line labeled with ``cut1'' in panel {\bf (d)} is used 
to investigate the longitudinal evolution of the jet whose time-slice diagram is displayed in 
Fig.~\ref{fig2}. The temporal evolution of jet1 is shown in a movie ({\it jet1.avi}) available 
in the online edition.}
\label{fig1}
\end{figure}

There were three recurrent and homologous EUV jets. The first jet (jet1) started at 
$\sim$21:29 UT and reached the maximum height around 21:41 UT
before falling back to the solar surface. Fig.~\ref{fig1} shows jet1 observed at 
$\sim$21:30:20 UT in six EUV filters. The jet was composed of a near-vertical collimated 
brightening that was 2.9 Mm in width at the top and a two-chamber structure at the bottom, 
which are commonplace in most inverse-$Y$-shaped coronal jets. However, a bright kernel 
visible in all the EUV wavelengths was ejected upwards along the jet as pointed by the white 
arrows.

\begin{figure}
\centering
\includegraphics[width=10cm,clip=]{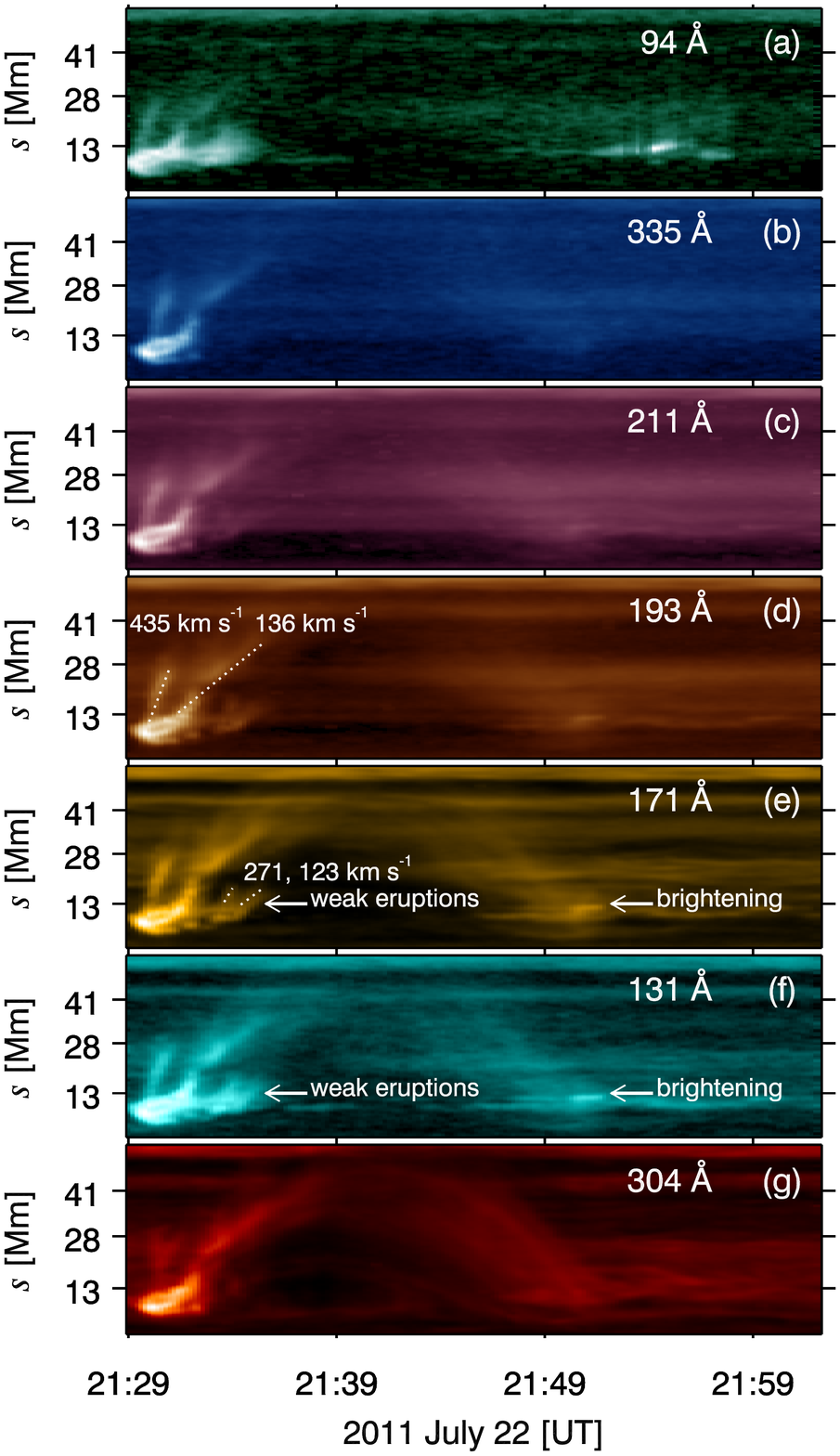}
\caption{Time-slice diagrams of cut1 in the seven filters during jet1. 
The white dotted lines in panel {\bf (d)} signify the
two major eruptions of the jet, with the slopes representing the rising speeds
of 435 km s$^{-1}$ and 136 km s$^{-1}$, respectively. The major eruptions
were followed by two weak and short eruptions illustrated by white dotted 
lines and pointed by the white arrows in panels {\bf (e)} and {\bf (f)}. The
slopes of the short dotted lines represent the velocities of the eruptions, being
271 km s$^{-1}$ and 123 km s$^{-1}$, respectively. The near-parabolic trajectory 
of the jet is more evident in the low-$T$ filters (131, 171, and 304
{\AA}) than in the high-$T$ filters (211, 193, and 335 {\AA}). Note the possible 
post-jet brightenings around 21:52 UT in panels {\bf (e)} and {\bf (f)}.}
\label{fig2}
\end{figure}

In order to study the longitudinal evolution of the jet, we extracted the intensity along the 
jet axis, which is labeled with ``cut1'' (75$\arcsec$ in length) and indicated by the white 
dashed line in Fig.~\ref{fig1}d. The time-slice 
diagrams of cut1 in the seven EUV wavelengths are displayed in Fig.~\ref{fig2}. It is
revealed that jet1 underwent two main eruptions at the speed of $\sim$435 
km s$^{-1}$ and $\sim$136 km s$^{-1}$ (Fig.~\ref{fig2}d), followed by two weak and 
short eruptions that are more obvious in 171 {\AA} (Fig.~\ref{fig2}e) and 131 {\AA} 
(Fig.~\ref{fig2}f). The velocities of the two weak short eruptions were 271 km s$^{-1}$
and 123 km s$^{-1}$. The length of the jet reached maximum (35.6 Mm) at $\sim$21:41 
UT before decreasing to zero when the jet fell back to the solar surface at $\sim$21:53 UT, 
which resulted in possible weak 
post-jet brightenings around 21:52 UT (see panels {\bf (e)} and {\bf (f)} in Fig.~\ref{fig2}). 
The near-parabolic trajectory and the falling phase of the jet are more evident in the low-$T$ 
filters (see panels {\bf (e)}, {\bf (f)}, and {\bf (g)} in Fig.~\ref{fig2}) than in the 
high-$T$ filters (see panels {\bf (b)}, {\bf (c)}, and {\bf (d)} in Fig.~\ref{fig2}).
Such trajectory, similar to the case reported by Zhang \& Ji (\cite{zqm14}), indicates
that the hot EUV jets underwent cooling owning to radiative loss and thermal conduction. 
The lifetime of jet1 is $\sim$24 min.

\begin{figure}
\centering
\includegraphics[width=10cm,clip=]{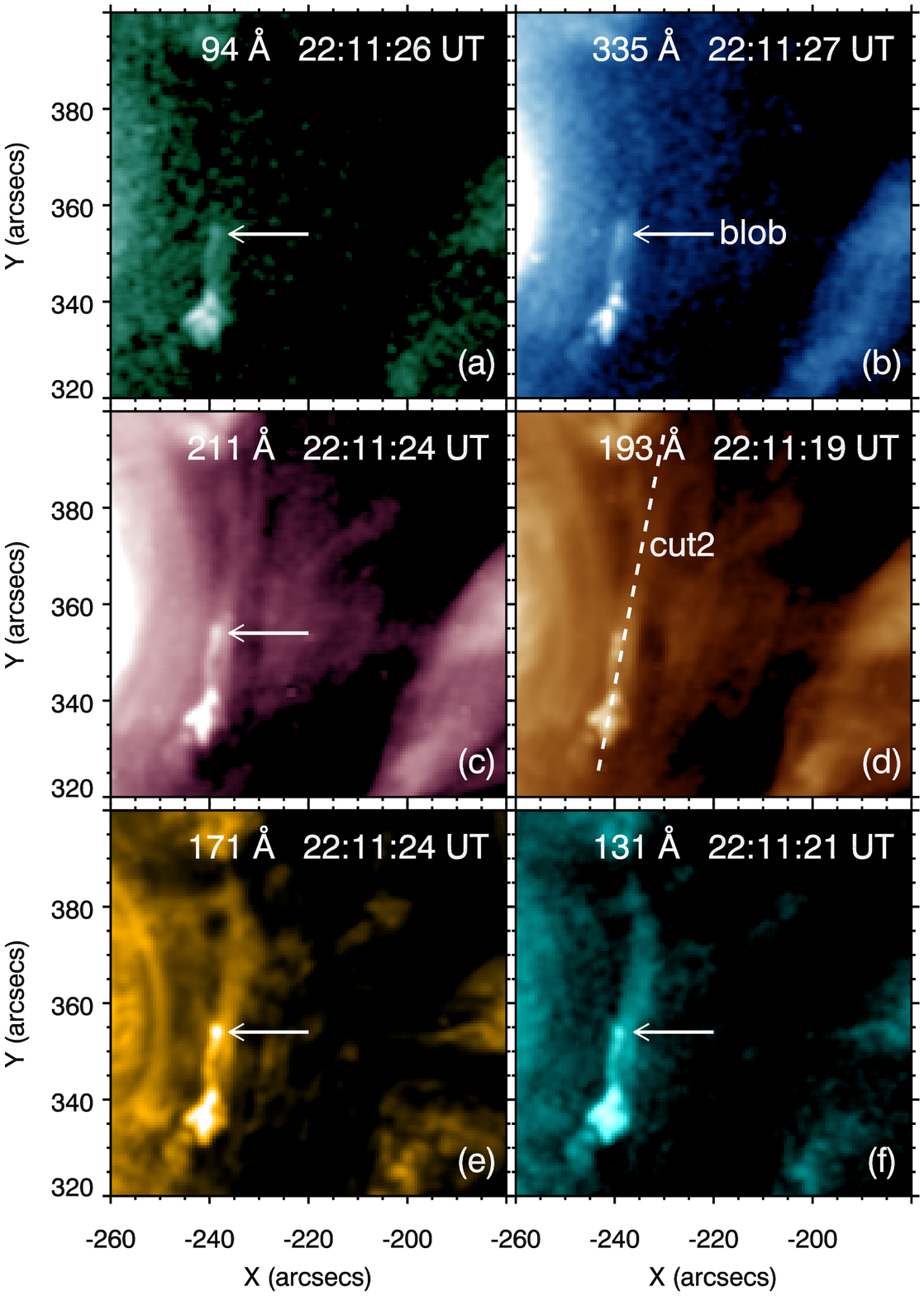}
\caption{Snapshots of jet2 seen in the six filters at $\sim$22:11:20 UT. 
The white arrows point to the blobs within the jet in panels {\bf (a)}, {\bf (b)}, 
{\bf (c)}, {\bf (e)}, and {\bf (f)}. The white dashed line labeled with ``cut2'' in panel {\bf (d)} 
is used to investigate the 
longitudinal evolution of the jet whose time-slice diagram is displayed in Fig.~\ref{fig4}.
The temporal evolution of jet2 is shown in a movie ({\it jet2.avi}) available in the 
online edition.}
\label{fig3}
\end{figure}

About a quarter after the end of jet1, the second collimated jet (jet2) occurred at the 
same place as the previous one. The bottom of the jet was composed of a couple of 
tiny bright kernels rather than a typical two-chamber structure.
It started at $\sim$22:10 UT and reached the maximum height at 
$\sim$22:19 UT before falling back to the solar surface at $\sim$22:34 UT.
Fig.~\ref{fig3} shows snapshots of the jet at $\sim$22:11:20 UT in six of the EUV filters.
Jet2 that was 5.4 Mm in width was more inclined and diffused than jet1. 
There was also a compact and bright feature within jet2, which is pointed
by the arrows.

\begin{figure}
\centering
\includegraphics[width=10cm,clip=]{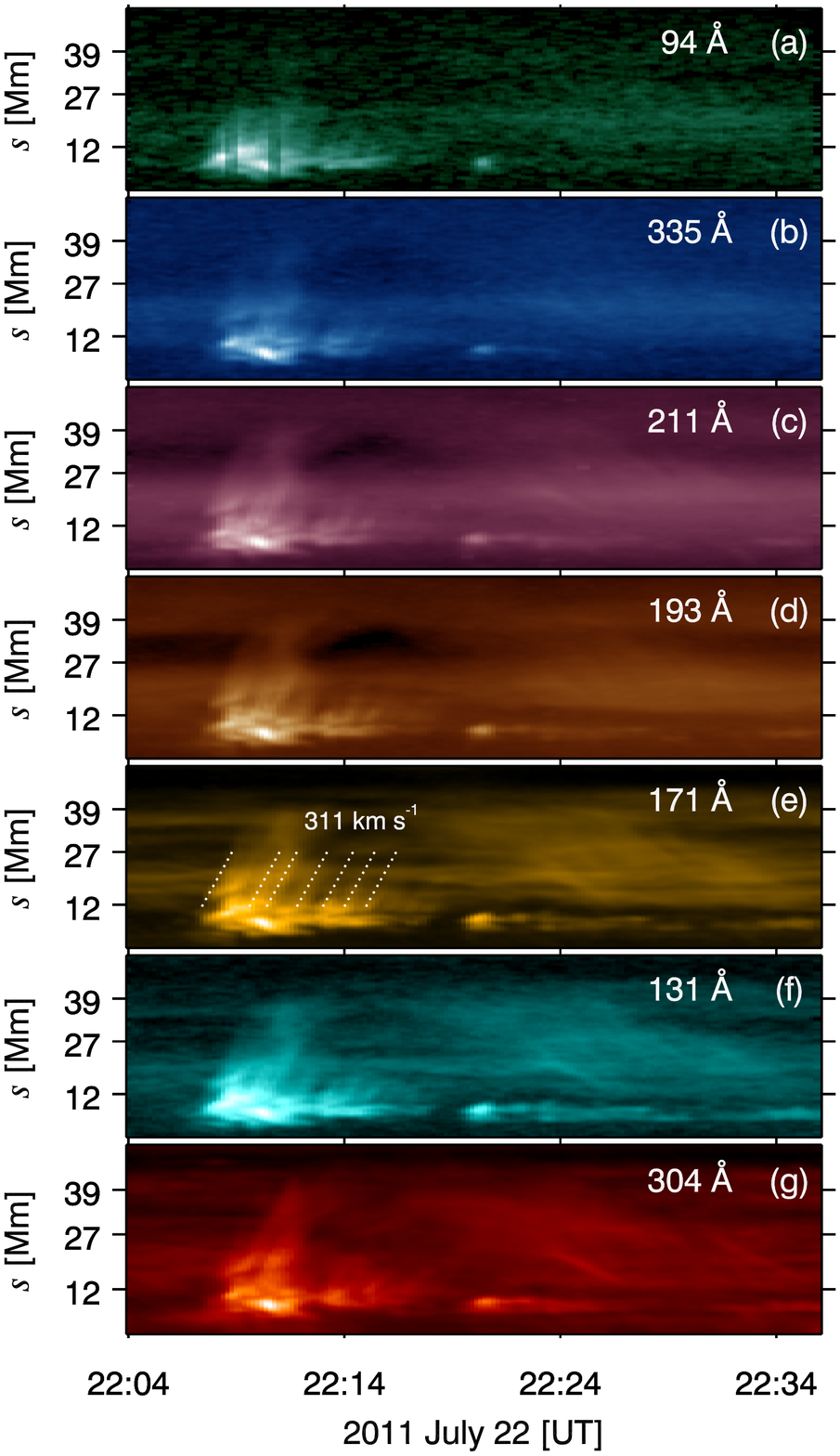}
\caption{Time-slice diagrams of cut2 in the seven filters during jet2. 
It experienced quasi-periodic upward eruptions 
with period of $\sim$65 s during 22:07$-$22:17 UT at the speed of $\sim$311 
km s$^{-1}$, which is indicated by the white dotted lines in panel {\bf (e)}.}
\label{fig4}
\end{figure}

We extracted the intensity along the axis of jet2, which is labeled with ``cut2''
(71$\arcsec$ in length) and indicated by the white dashed line in Fig.~\ref{fig3}d. The
time-slice diagrams of cut2 in the seven filters are shown in Fig.~\ref{fig4}.
Starting at 22:07 UT, jet2 experienced a series of quasi-periodic eruptions 
with period of $\sim$65 s until 22:17 UT at the speed of $\sim$311 km s$^{-1}$, which 
are illustrated by the multiple white dashed dotted lines in Fig.~\ref{fig4}e. 
Like jet1, jet2 also presented near-parabolic trajectory, 
i.e., plasma fell back to the solar surface after reaching the maximum height (31.7 Mm). 
The falling phase that ended at 22:35 UT was most distinct in the low-$T$ wavelengths 
(see panels {\bf (e)}, {\bf (f)}, and {\bf (g)}).

\begin{figure}
\centering
\includegraphics[width=10cm,clip=]{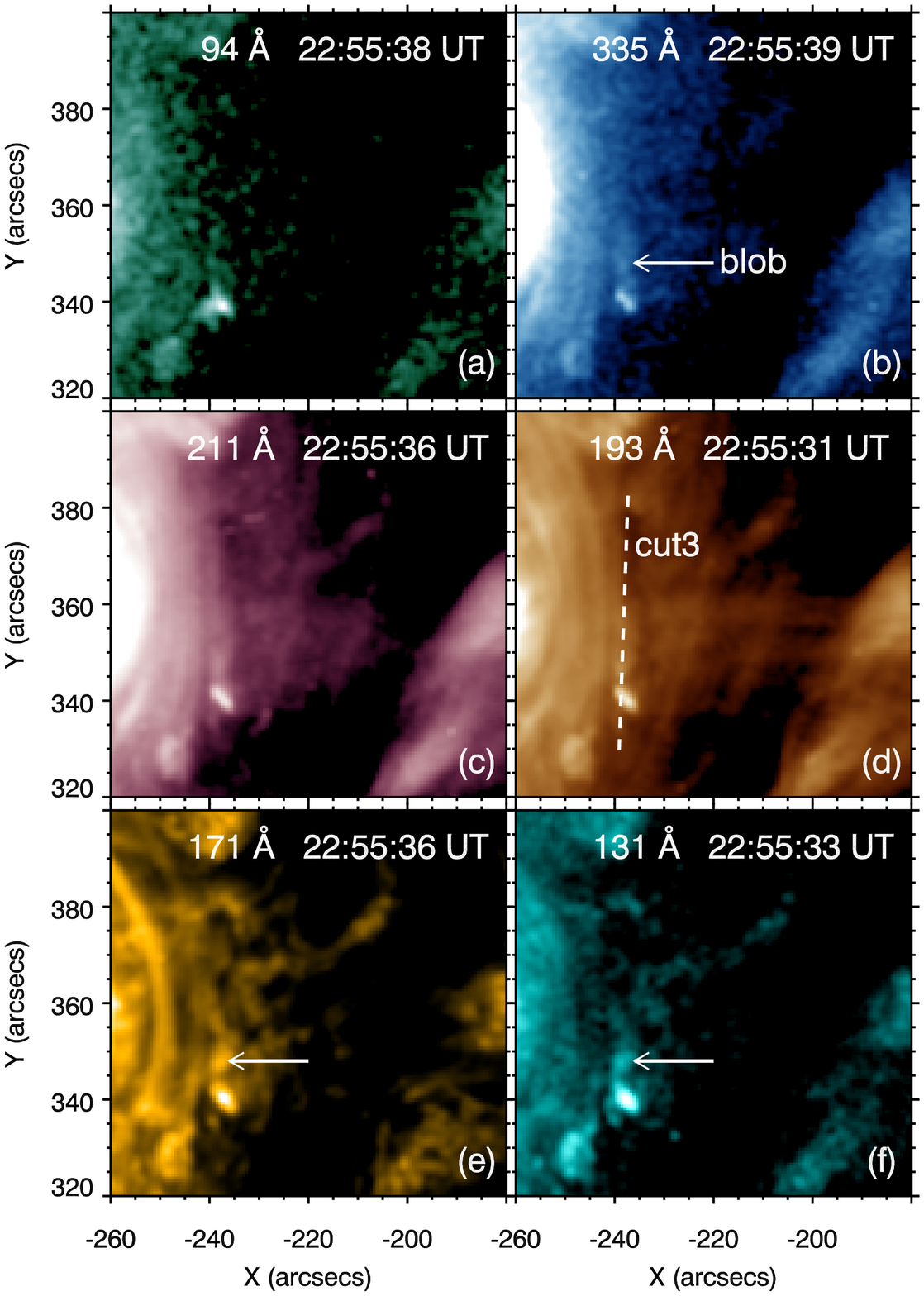}
\caption{Snapshots of jet3 seen in the six filters at $\sim$22:55:35 UT. 
The white arrows point to the blobs within the jet in panels {\bf (b)}, {\bf (e)}, 
and {\bf (f)}. The white dashed line labeled with ``cut3'' in panel {\bf (d)} is used 
to investigate the longitudinal evolution of the jet whose time-slice diagram is 
displayed in Fig.~\ref{fig6}. The temporal evolution of jet3 is shown in a 
movie ({\it jet3.avi}) available in the online edition.}
\label{fig5}
\end{figure}

\begin{figure}
\centering
\includegraphics[width=10cm,clip=]{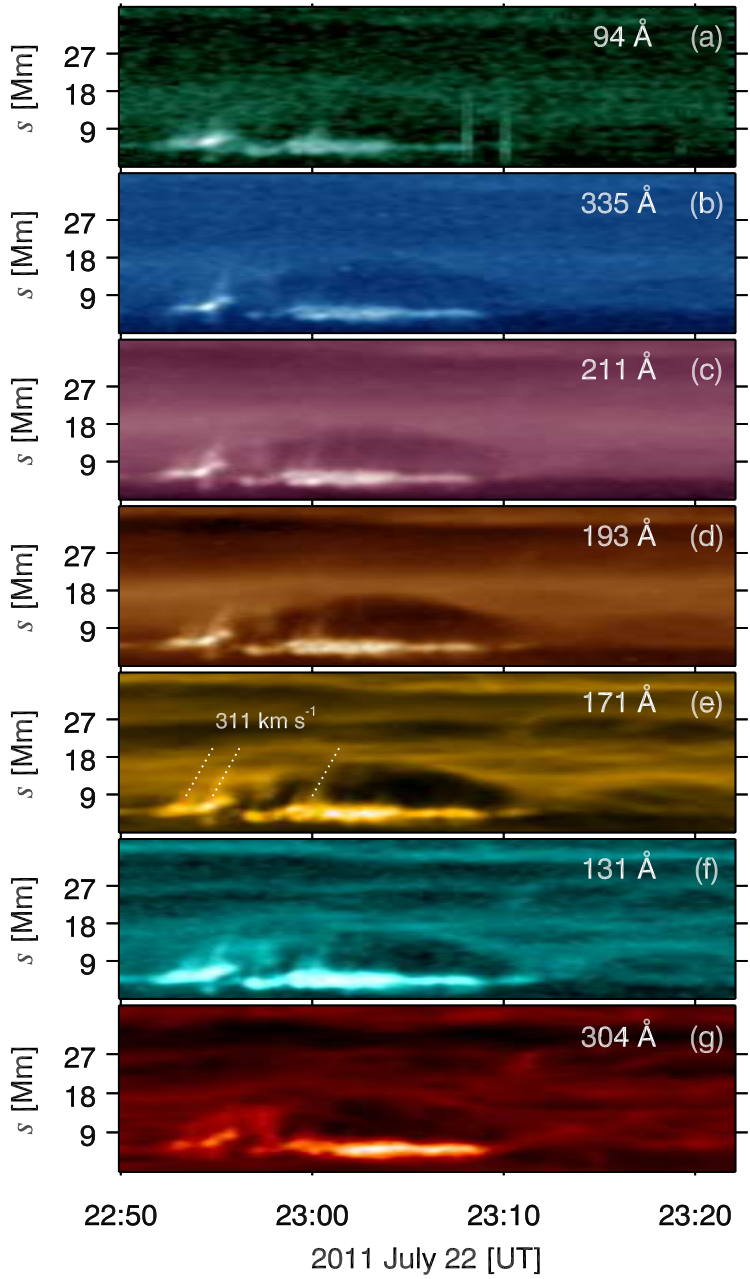}
\caption{Time-slice diagrams of cut3 in the seven filters during jet3. 
The white dotted lines in panel {\bf (e)} illustrate
the intermittent eruptions of jet3 during 22:52$-$23:02 UT. The slopes
of the dotted lines stand for the rising velocities of the eruptions, i.e., 311 
km s$^{-1}$.}
\label{fig6}
\end{figure}

After 22:52 UT, the third homologous jet (jet3) appeared, but with much shorter length and 
weaker intensity compared with the previous jets. Fig.~\ref{fig5} shows snapshots of 
the jet in the six filters at $\sim$22:55:35 UT. The bright and compact feature of jet3 
is more evident in the 131, 171, and 335 {\AA} images. We 
extracted the intensity along its axis, which is labeled with ``cut3'' (53$\arcsec$ in length) 
and indicated by the white dashed line in Fig.~\ref{fig5}d. The time-slice diagrams of cut3
in the seven filters are displayed in Fig.~\ref{fig6}. Like the previous jets, jet3 underwent 
intermittent eruptions, which are denoted by the white dotted lines 
in Fig.~\ref{fig6}e, and reached maximum height ($\sim$11.2 Mm) before falling back to 
the solar surface. The rising velocities of the eruptions were $\sim$311 km s$^{-1}$. The 
near-parabolic trajectory of jet3 is clearly illustrated in the 131, 193, 171, and 304 
{\AA} images. The parameters of the three EUV jets are summarised in Table~\ref{table:1}, 
including the maximum apparent heights, widths, apparent rising velocities, and lifetimes.

\begin{table}
\caption{Parameters of the three EUV jets} 
\label{table:1}
\centering
\begin{tabular}{l c c c c}
\hline\hline
No. & Height & Width & Velocity & Lifetime \\ 
       &  (Mm)  &  (Mm)  & (km s$^{-1}$) & (min)  \\
\hline
  1 & 26.8 & 2.9 & 123$-$435 & 24 \\
  2 & 23.7 & 5.4 & $\sim$311 & 28 \\
  3 & 12.8 & 2.8 & $\sim$311 & 20 \\
\hline
\end{tabular}
\end{table}

\subsection{Blobs in the jets} \label{s-plas}

\begin{figure}
\centering
\includegraphics[width=10cm,clip=]{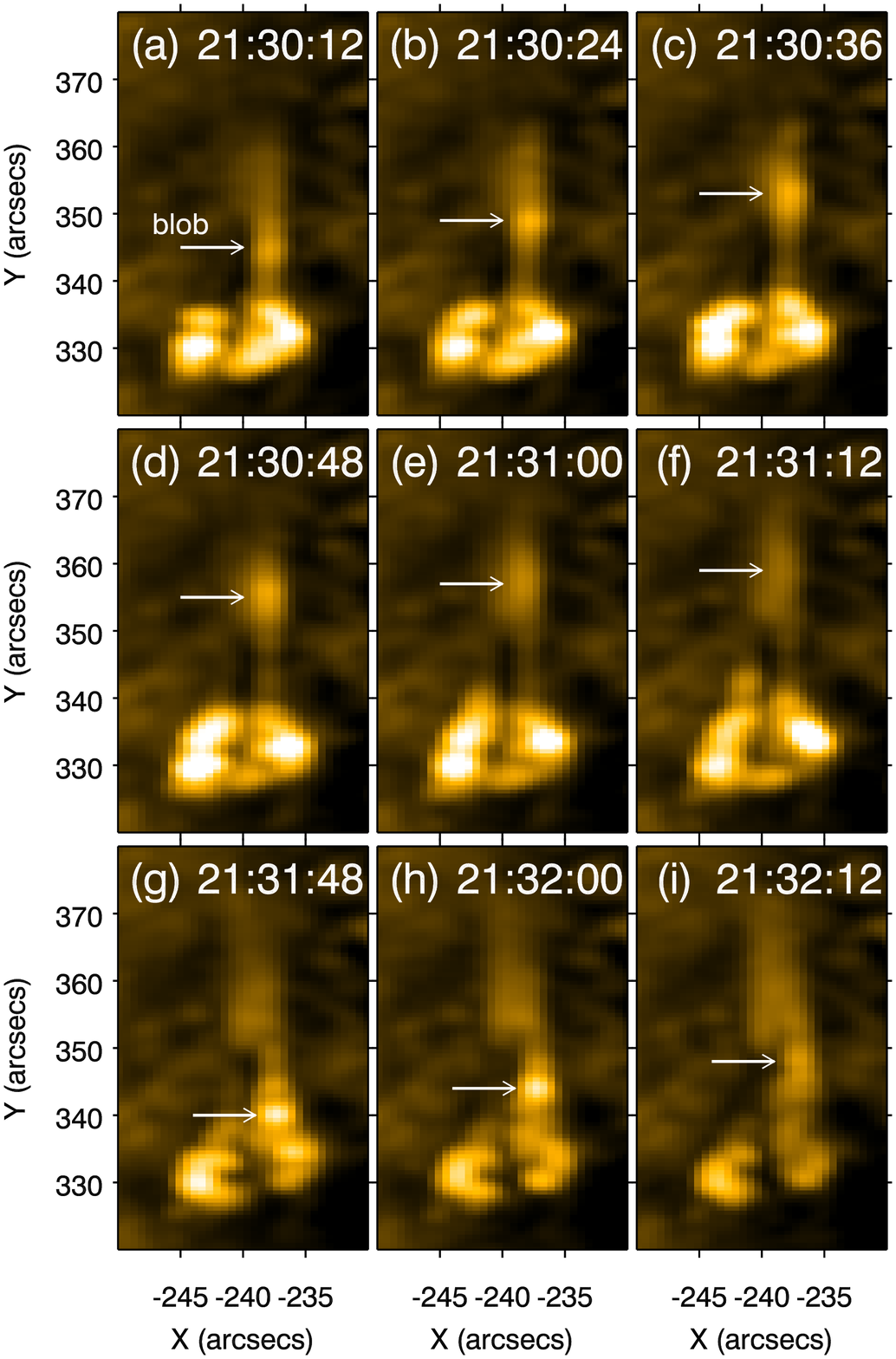}
\caption{{\bf (a)}$-${\bf (i)} Nine snapshots of the AIA 171 {\AA} images, 
showing the blobs during jet1 as pointed by the white arrows.}
\label{fig7}
\end{figure}

From the online movies that illustrate the evolutions of the recurrent jets, we identified 
more blobs. Fig.~\ref{fig7} shows nine snapshots of the 171 {\AA} images during 
jet1. The bright and compact kernels as pointed by the white arrows moved 
upwards from the bottom to the top of the jet during 21:30:12$-$21:31:12 UT and 
became blurred after mixing with the surroundings. After careful inspection, we found 
that the size of blob increased slightly during the ejection. About 30 s later, another 
blob appeared at the bottom and rose until 21:32:12 UT. The sizes of the blobs 
were $\sim$3 Mm. The intermittent appearances, 
upwards ejections, and disappearances of blobs were consistent with the sporadic 
eruptions of the jet, implying the bursty nature of magnetic reconnection in the jet.

\begin{figure}
\centering
\includegraphics[width=10cm,clip=]{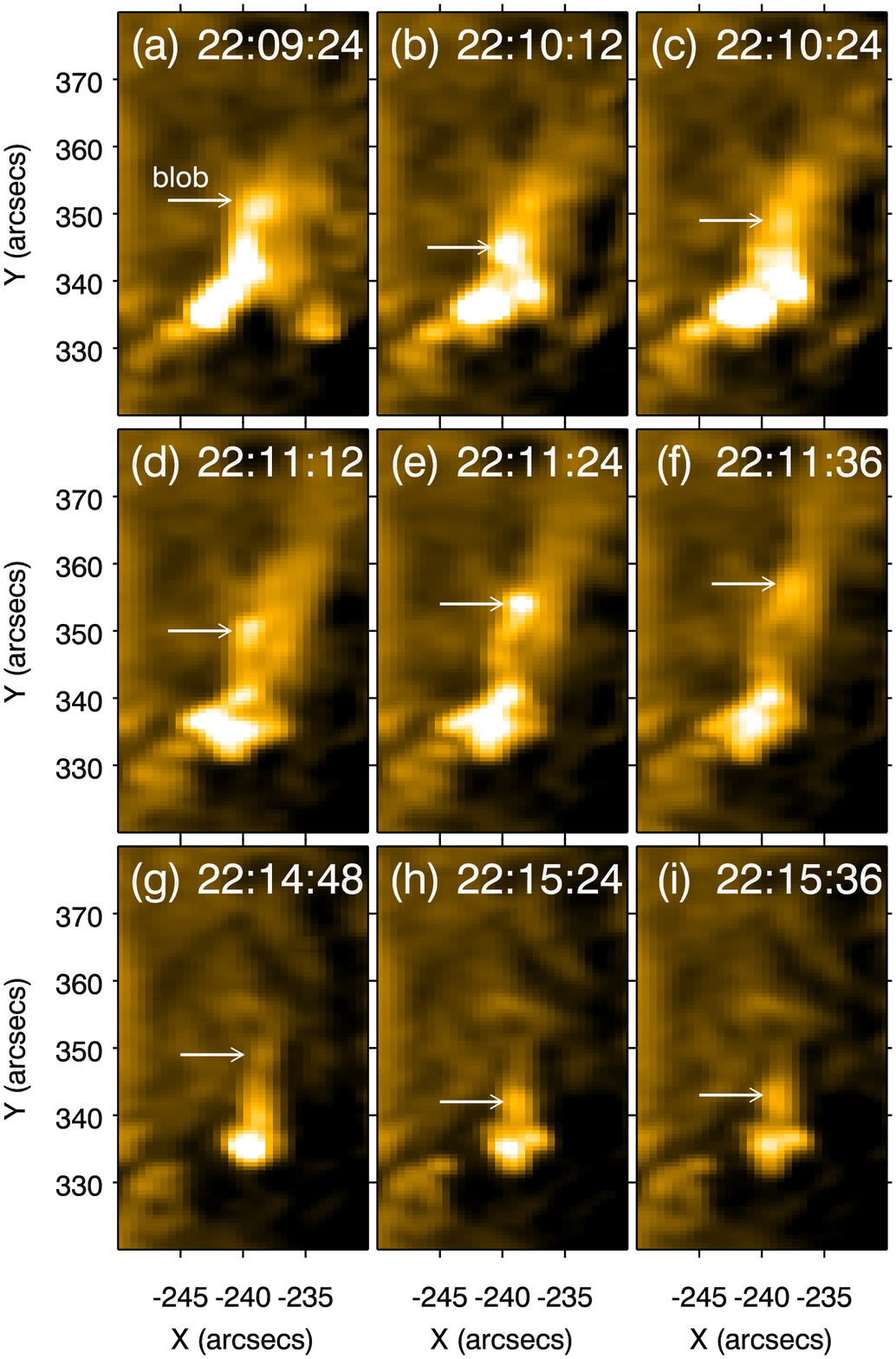}
\caption{{\bf (a)}$-${\bf (i)} Nine snapshots of the AIA 171 {\AA} images, 
showing the blobs during jet2 as pointed by the white arrows.}
\label{fig8}
\end{figure}

Figure~\ref{fig8} displays nine snapshots of the 171 {\AA} images during jet2 
within which recurrent blobs were ejected. Despite of the high cadence of AIA (12 s), 
a blob was clearly present in 2$-$5 EUV snapshots due to their short lifetimes (24$-$60 s). 
Panels {\bf (d)}$-${\bf (f)} illustrate the rising motion of a blob from the middle 
to top of the jet. Panels {\bf (h)}$-${\bf (i)} show a blob that was restricted near the 
bottom of the jet. 

\begin{figure}
\centering
\includegraphics[width=10cm,clip=]{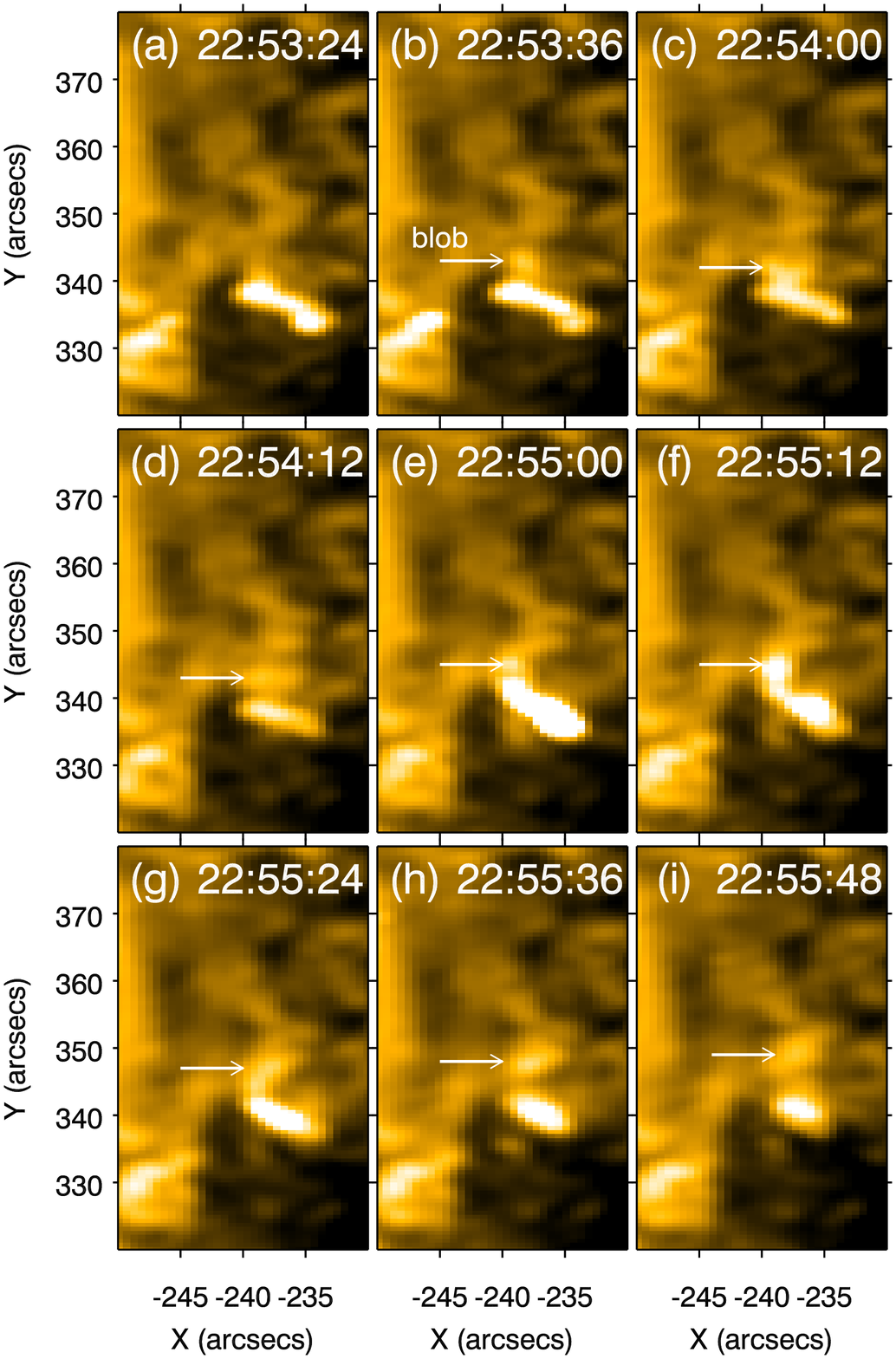}
\caption{{\bf (a)}$-${\bf (i)} Nine snapshots of the AIA 171 {\AA} images, 
showing the blobs during jet3 as pointed by the white arrows.}
\label{fig9}
\end{figure}

Figure~\ref{fig9} displays nine snapshots of the 171 {\AA} images during jet3. 
Compared with the previous two jets, the blobs were much fainter and reached 
lower heights. Panels {\bf (b)}$-${\bf (d)} reveal that the blobs were close to the 
bottom of jet. Panels {\bf (e)}$-${\bf (i)} illustrate the complete evolution of a blob. 
It appeared at 22:55:00 UT and reached maximum brightness at 22:55:12 UT 
before gradually mixing with the surroundings and fading out after 22:55:48 UT.

\begin{figure}
\centering
\includegraphics[width=8cm,clip=]{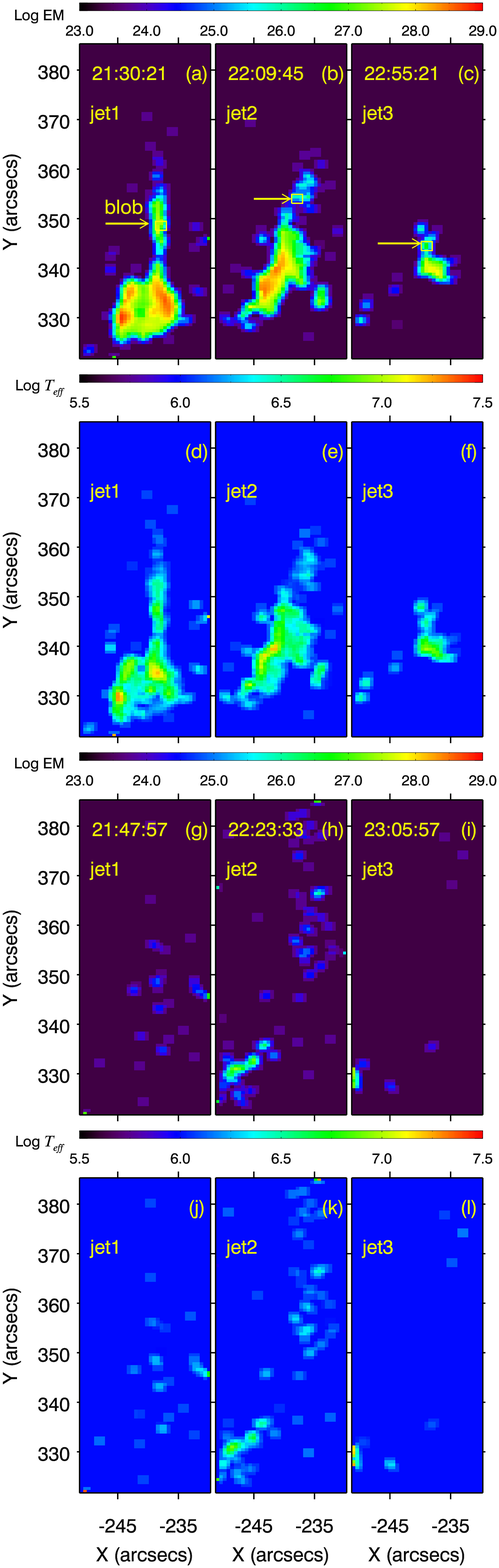}
\caption{{\it Top two rows}: EM {\bf (a$-$c)} and temperature maps {\bf (d$-$f)} of the 
jets during their rising phases. The arrows point to the blobs in the jets. The DEM curves 
of the plasmas in the boxes of {\bf (a$-$c)} are displayed in Fig.~\ref{fig11}. 
{\it Bottom two rows}: EM {\bf (g$-$i)} and temperature maps {\bf (j$-$l)} of the jets during 
their falling phases. Note that EM and $T_{eff}$ are in $\log$-scales.}
\label{fig10}
\end{figure}

We performed DEM analyses and derived the two-dimensional (2D) distributions 
of the EM and temperature ($T_{eff}$) of the jets. Note that the minimum and maximum 
temperatures ($T_1$ and $T_2$) for the integral of EM are 10$^{5.5}$ K and 10$^{7.5}$ 
K. In Fig.~\ref{fig10},
the top two rows demonstrate the selected EM and $T_{eff}$ maps of the jets 
during the rising phases of jet1 ({\it left}), jet2 ({\it middle}), and jet3 ({\it right}), 
respectively. The jets and blobs pointed by the yellow arrows are clearly present in the 
maps with higher emissions and temperatures than the adjacent quiet region.
The two-chamber base of jet1 and the cusp-like bases of jet2 and jet3 with the highest 
emissions and temperatures are also clearly demonstrated in the maps.
The bottom two rows demonstrate the selected EM and $T_{eff}$ maps of the jets 
during the falling phases of jet1 ({\it left}), jet2 ({\it middle}), and jet3 ({\it right}), 
respectively. It is seen that the emissions of the jets got quite weak and the temperatures 
decreased to a low level close to the adjacent quiet region, which is consistent with the 
time-slice diagrams of the jets in Figs.~\ref{fig2}, \ref{fig4}, and \ref{fig6}.

\begin{figure}
\centering
\includegraphics[width=10cm,clip=]{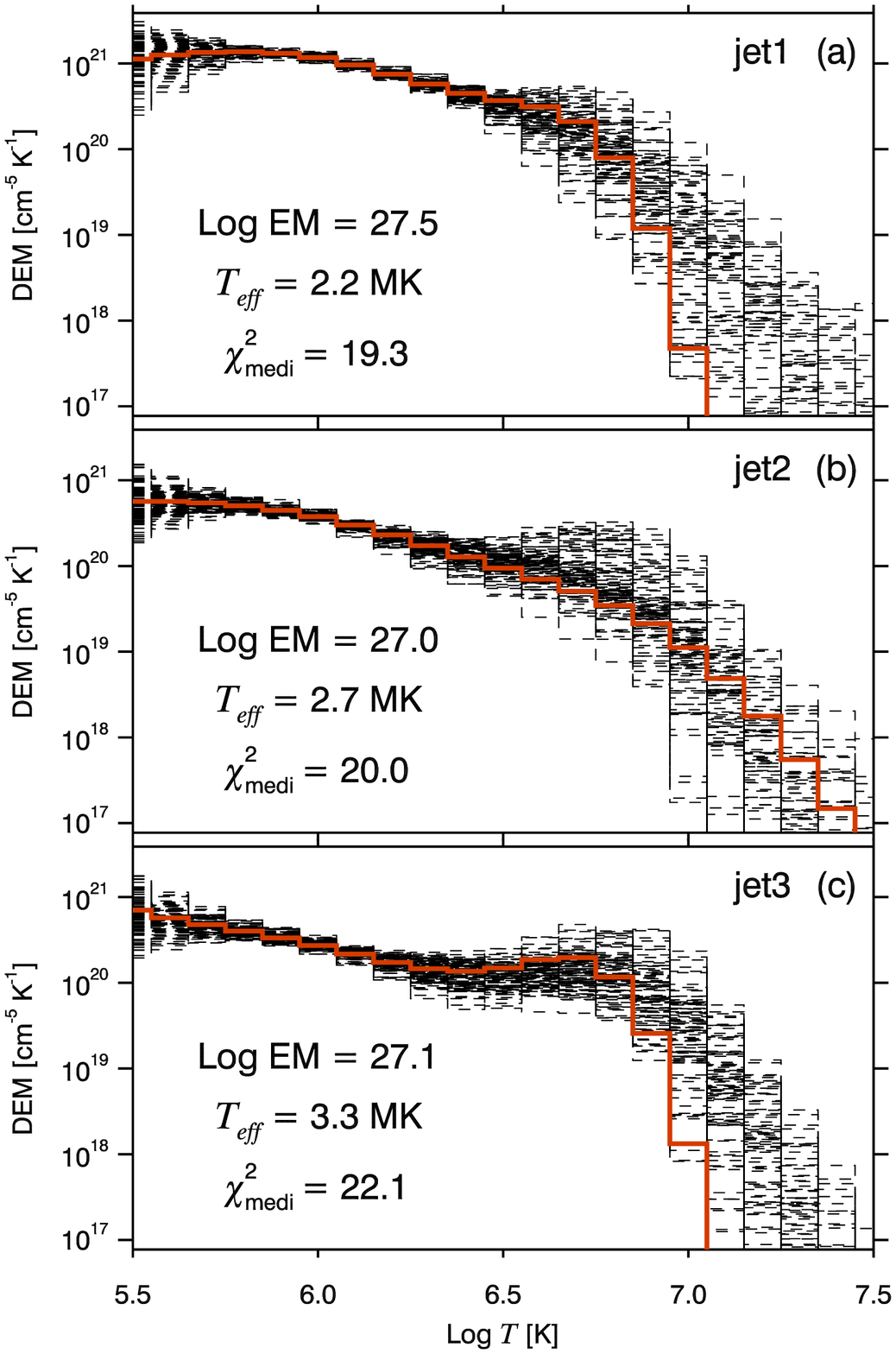}
\caption{DEM profiles of the blob core regions of jet1 {\bf (a)}, jet2 {\bf (b)}, 
and jet3 {\bf (c)} indicated in the top panels of Fig.~\ref{fig10}. The red solid lines 
stand for the best-fitted DEM curves from the observed values. The black dashed 
lines represent the reconstructed curves from the 100 MC simulations. The 
corresponding EM, $T_{eff}$ of the blobs, and the median values of $\chi^2$ 
of the MC simulations are displayed.}
\label{fig11}
\end{figure}

In the top panels of Fig.~\ref{fig10}, the core regions of blobs are included in the 
small yellow boxes with sizes of 1.8$\arcsec$. The base-difference EUV intensities within 
the boxes were averaged so that the core regions were taken as a whole. The DEM profiles 
of the first, second, and third blob cores are displayed in the top, middle, and bottom panels of 
Fig.~\ref{fig11}, respectively. The red solid lines stand for the best-fitted profiles derived from 
the observed values, while the black dashed lines represent the profiles derived from the 100 
MC simulations. It is obvious that the DEM profiles have a broad range between 
$5.5\le \log T \le 7.5$, indicating that the blobs are multi-thermal in nature. However, the 
contributions of EM come mainly from the low-$T$ plasma since the DEM decreases with 
$\log T$. The reconstructions of the curves are most accurate and reliable in the range of 
$5.5\le \log T \le 6.5$. The scatter increases with temperature in the range of 
$6.5\le \log T \le 7.5$. The median values of the $\chi^2$ in the orders of 20 are labeled 
in Fig.~\ref{fig11}. The $T_{eff}$ of the blob core regions of jet1, jet2, and jet3 are 2.2, 2.7, 
and 3.3 MK. The corresponding $\log \mathrm{EM}$ are 27.5, 27.0, and 27.1, respectively. 
The average 
electron number densities of the blobs $n_e$ were estimated according to $\sqrt{\mathrm{EM/H}}$ 
(assuming filling factor $\approx$1), where EM and H stand for the EM and LOS depth of the 
blobs. Assuming that the jets are cylindric, the LOS depths equalled to the apparent widths of 
the blobs. The values of $n_e$ for the three blobs are 3.3, 1.9, and 2.1$\times$10$^9$ cm$^{-3}$, 
respectively. However, such estimations using the imaging data are very qualitative due to the 
large uncertainties of the plasma filling factor as well as the LOS column depth of the blobs.
Considering the favorable perspectives of the Solar Terrestrial Relation Observatory (STEREO; 
Kaiser et al. \cite{kai05}), we tried to find the counterparts of the jets in the EUV images observed by 
STEREO but failed due to the small scales and weak intensities of the jets. More precise diagnostics 
of the plasma densities should be conducted using the spectra-imaging observations.

We also derived the DEM curves of the blobs in jet1, jet2, and jet3 during their lifetimes. 
They are similar to those displayed in Fig.~\ref{fig11}, featuring broad distributions and decreasing 
trends with temperature. Base on the DEM curves, we calculated the temperatures of the blobs in 
the three recurrent jets. The $T_{eff}$ ranges from 0.5 to 4 MK with an median value of 2.3 MK.

\section{Discussion} \label{s-disc}

\subsection{Which type do the recurrent jet belong to?} \label{s-type}

Despite of their small scales, coronal jets usually present various morphology and 
characteristics. Nistic{\`o} et al. (\cite{nis09}) classified the 79 polar jets into four 
catalogues: Eiffel Tower-type jets, $\lambda$-type jets, micro-CME-type, and others. 
Based on the physical mechanisms, Moore et al. (\cite{moo10}) classified polar jets into 
standard and blowout jets. The former are the same as the well-known inverse-$Y$ jets.
The latter are counterparts of erupting-loop H$\alpha$ macrospicules, where the 
jet-base magnetic arch undergoes a miniature version of the blowout eruptions that 
produce major CMEs. The main differences lie in whether the base arches have  
enough shear and twist to erupt open. 
Moore et al. (\cite{moo13}) carried out in-depth comparison and found that the blowout 
jets statistically have cool component seen in 304 {\AA}, lateral expansion, and axial rotation.
The recurrent jets in our study matched the standard type according to their morphology. 
However, they were present in all the EUV wavelengths of AIA, including the cool filters 
(304 {\AA}). Lateral expansion and axial rotation, however, were absent in the jets. We 
have not noticed signatures of CME-like eruptions from the jet bases or curtain-like 
shapes after their eruptions as in the blowout jets. Therefore, we propose that the recurrent 
jets belong to the standard type though they had cool component.

\subsection{Relationship between jets and surges} \label{s-rel}

\begin{figure}
\centering
\includegraphics[width=10cm,clip=]{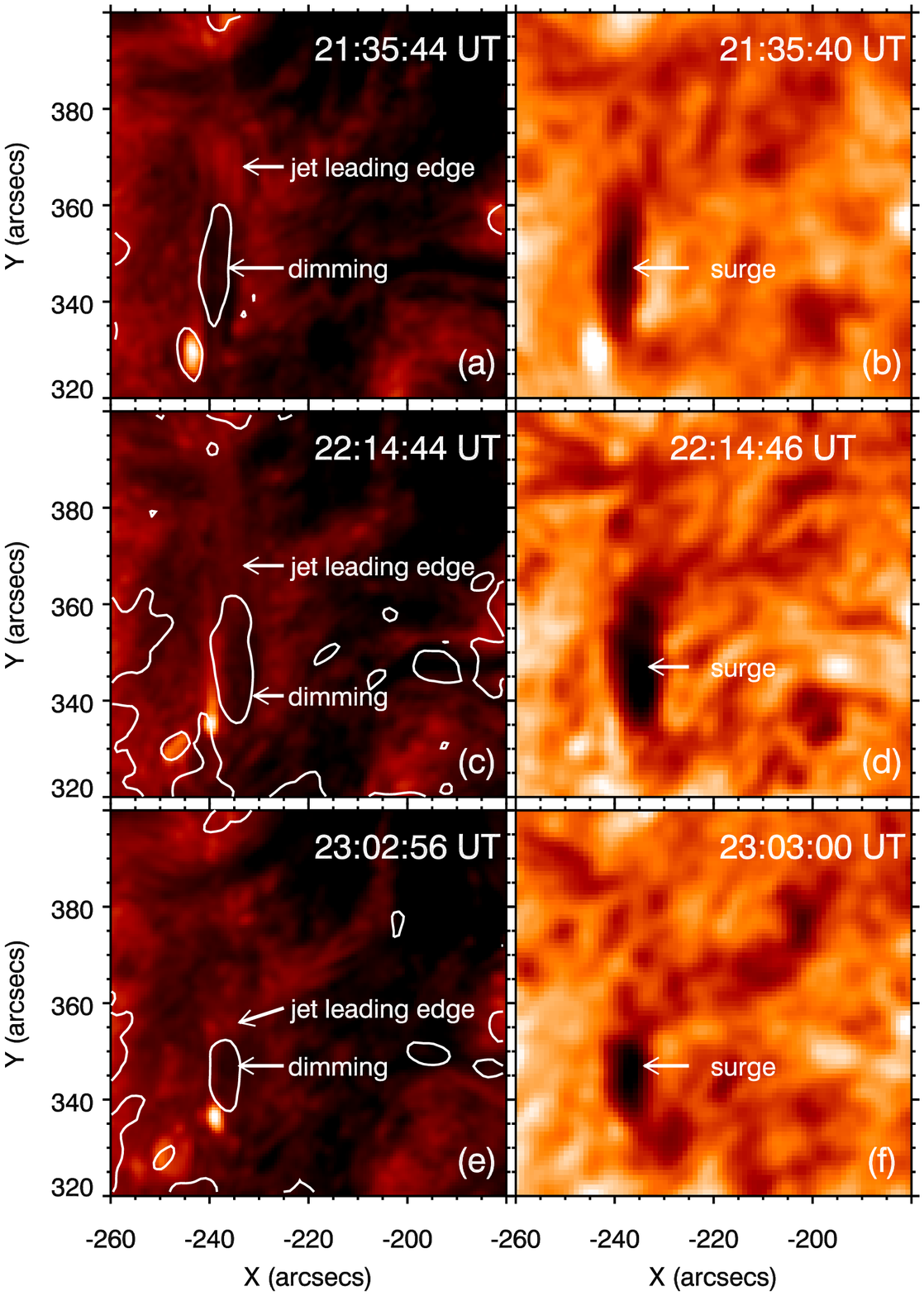}
\caption{{\it Left panels}: AIA 304 {\AA} images during jet1 {\bf (a)}, 
jet2 {\bf (c)}, and jet3 {\bf (e)}. The arrows point to the leading edges 
of the jets and the following dimming regions. {\it Right panels}: 
The near-simultaneous H$\alpha$ images of the same FOV. 
The intensity contours of the H$\alpha$ images are overlaid on the 
corresponding 304 {\AA} images.}
\label{fig12}
\end{figure}

Coronal jets are always observed in the EUV and X-ray wavelengths, while surges are often 
observed in H$\alpha$ due to their cool nature. According to the numerical simulations of 
Yokoyama \& Shibata (\cite{yoko96}), 
both hot (10$^5$$-$10$^7$ K) and cool ($\sim$10$^4$ K) plasma ejections are created 
side by side during the magnetic reconnection between the emerging flux and the pre-existing 
magnetic fields. The spatial relationship between the jets and surges is controversial. It has 
been observed that surges are adjacent to jets (Canfield et al. \cite{can96}; Chae et al. \cite{chae99}; 
Jiang et al. \cite{jiang07}). In our study, the hot jets observed in the AIA filters were 
associated with recurrent cool surges observed and covered during their whole lifetimes in 
the H$\alpha$ line center (Wang et al. \cite{wang14}). In Fig.~\ref{fig12}, we 
compared the 304 {\AA} ({\it left panels}) and H$\alpha$ ({\it right panels}) images that 
represent the three jets from top to bottom rows, respectively. We also superpose
the intensity contours of the H$\alpha$ images on the corresponding EUV images,
finding that the surges were cospatial with the EUV dimmings behind the leading 
edges of the jets. After examining the movies of the recurrent jets in the other six
EUV filters, we found that the dimmings visible in all the filters were 
cospatial with the surges. In the swirling flare-related jet on 2011 October 15 at the edge 
of AR 11314,  Zhang \& Ji (\cite{zqm14}) discovered EUV dimming behind the leading
edge of the jet, which was explained by the absorption of the EUV emissions of 
the hot jet by the foreground cool surge. Such explanation could be convincingly 
justified by our detailed and complete case study of recurrent jets. Considering that 
the jets and surges are 3D in nature (Moreno-Insertis \& Galsgaard \cite{mor13}), 
we believe that the differences of spatial relationship between the jet and surge lie in the 
different perspectives of observation. When the LOS is perpendicular to the plane determined 
by the jet and surge, they are adjacent to each other. However, when the LOS is parallel to the 
plane, they are cospatial and the EUV and X-ray emissions of the jets may be absorbed by the 
surges.

\subsection{Physical properties of the blobs} \label{s-pro}

The magnetic islands or plasmoids associated with solar flares have extensively been 
observed and studied. Asai et al. (\cite{asa04}) discovered bursty sunward motions above 
the post-flare loops. The authors interpreted the bursty downflow as many plasmoids created inside 
the current sheet induced by the eruption of the large-scale filament. The velocities of the downflow 
were 45$-$500 km s$^{-1}$, the electron number densities were 1$-$10$\times$10$^9$ cm$^{-3}$, 
and the sizes were 2$-$10 Mm. In our case, the rising velocities of the blobs in the jets were 
120$-$450 km s$^{-1}$, which were close to the values of the plasmoids associated with the big flare.
The sizes ($\sim$3 Mm) of the blobs in the jets were also comparable to those of the flare-related 
plasmoids. Using the AIA multi-wavelength observations and the method of DEM reconstruction 
developed by Aschwanden et al. (\cite{asch13}), Kumar \& Cho (\cite{kum13}) studied the bidirectional 
plasmoid ejections whose DEM profiles have broad distributions, which are consistent with the cases 
of jet blobs in Fig.~\ref{fig11}, implying the multi-thermal nature of the blobs. Based on the above 
comparison, we propose that the physical properties (temperature, velocity, and size) of the blobs 
intermittently ejected out of the recurring jets are quite close to those of the ejected plasmoids resulting 
from the tearing-mode instability of the current sheets where magnetic reconnections take place during 
big flares.

\section{Summary} \label{s-sum}

In this paper, we studied the recurrent and homologous jets that took place at the western edge of AR 
11259 on 2011 July 22. The jets that had lifetimes of 20$-$30 min recurred for three times with interval 
of 40$-$45 min. Quasi-periodic eruptions were observed during each of the jets at the speed of 
120$-$450 km s$^{-1}$. After reaching the maximum heights, the jet plasmas returned back to the solar 
surface, showing near-parabolic trajectories. The falling phases were more evident in the low-$T$ filters 
than in the high-$T$ filters, indicating that the jets experienced cooling due to radiative loss and thermal 
conduction after the onset of eruptions. We identified very bright and compact features, i.e., blobs, in the 
jets during their rising phases. The simultaneous presences of blobs in all the EUV filters are consistent 
with the broad ranges of the DEM profiles of the blobs, indicating their multi-thermal nature. The DEM-weighted 
average temperatures of the blobs range from 0.5 to 4 MK with a median value of $\sim$2.3 MK.
The lifetimes of the blobs were 24$-$60 s. To our knowledge, this is the first report of blobs in coronal jets, 
and their physical properties are quite close to those of the bidirectionally ejected plasmoids during big 
flares, suggesting that the basic process of tearing-mode 
instability in current sheets exists not only in the large-scale solar flares but also in small-scale jets. 
Additional case studies and numerical simulations are required to get a better understanding of the blobs.

\begin{acknowledgements}
The authors are grateful to the referee for the enlightening and valuable comments.
Q.M.Z acknowledges X. Cheng, T. H. Zhou, Y. N. Su, B. Kliem, L. Ni, P. F. Chen, 
M. D. Ding, C. Fang, R. Moore, E. Pariat, and the solar physics group in 
Purple Mountain Observatory for discussions and suggestions. SDO is a mission 
of NASA\rq{}s Living With a Star Program. AIA and HMI data are courtesy of the 
NASA/SDO science teams. The H$\alpha$ data were obtained from the Global 
High Resolution H$\alpha$ Network operated by the Big Bear Solar Observatory, 
New Jersey Institute of Technology. This work is supported by 973 program under 
grant 2011CB811402 and by NSFC 11303101, 11333009, 11173062 and 11221063.
\end{acknowledgements}

\end{document}